\documentclass[sigconf,authorversion,nonacm]{acmart}
\usepackage{enumitem}

\usepackage{amsmath,amsfonts}
\usepackage{algorithmic}
\usepackage{algorithm}
\usepackage{graphicx}
\usepackage{textcomp}
\usepackage{threeparttable}
\usepackage{xcolor}
\usepackage{multirow}
\usepackage{pifont}
\usepackage{hyperref}
\hypersetup{colorlinks,linkcolor={blue},citecolor={blue},urlcolor={red}}  
\usepackage{bbding}
\usepackage{tikz}
\newcommand*\circled[1]{\tikz[baseline=(char.base)]{
            \node[shape=circle,draw,inner sep=1.1pt] (char) {#1};}}
\usepackage{listings}
\usepackage{xcolor}
\usepackage{framed}
\newcommand{\cmark}{\ding{51}}%
\newcommand{\xmark}{\ding{55}}%

\usepackage[strict]{changepage}

\usepackage{framed}
\usepackage{dblfloatfix}

\newcommand{\blzs}{\textsuperscript{$\blacklozenge$}}

\definecolor{formalshade}{rgb}{0.95,0.95,0.95}
\definecolor{ballblue}{rgb}{0.13, 0.67, 0.8}
\newenvironment{formal}{%
  \MakeFramed{\advance\hsize-\width\FrameRestore}%
  \noindent\hspace{-4.55pt}% disable indenting first paragraph
  \begin{adjustwidth}{}{7pt}%
  \vspace{2pt}\vspace{2pt}%
}
{%
  \vspace{2pt}\end{adjustwidth}\endMakeFramed%
}

\newenvironment{formal1}{%
  \MakeFramed{\advance\hsize-\width\FrameRestore}%
  \noindent\hspace{-4.55pt}% disable indenting first paragraph
  \begin{adjustwidth}{}{7pt}%
  \vspace{2pt}\vspace{2pt}%
}
{%
  \vspace{2pt}\end{adjustwidth}\endMakeFramed%
}

\newcommand{\synfuzz}{{\sffamily\textbf{SynFuzz}}}

\definecolor{vgreen}{RGB}{104,180,104}
\definecolor{vblue}{RGB}{49,49,255}
\definecolor{vorange}{RGB}{255,143,102}

\lstdefinestyle{verilog-style}
{
    language=Verilog,
    basicstyle=\small\ttfamily,
    keywordstyle=\color{vblue},
    identifierstyle=\color{black},
    commentstyle=\color{vgreen},
    numbers=left,
    numberstyle=\tiny\color{black},
    numbersep=10pt,
    tabsize=8,
    moredelim=*[s][\colorIndex]{[}{]},
    literate=*{:}{:}1
}

\def\BibTeX{{\rm B\kern-.05em{\sc i\kern-.025em b}\kern-.08em
    T\kern-.1667em\lower.7ex\hbox{E}\kern-.125emX}}

\AtBeginDocument{%
  \providecommand\BibTeX{{%
    Bib\TeX}}}

\begin{document}


% \title{SynFuzz: Fuzzing of Netlist to Detect Synthesis Bugs                
% }

\title{SynFuzz: Leveraging Fuzzing of Netlist to Detect Synthesis Bugs           
}

%%\orcid{1234-5678-9012}
%author{G.K.M. Tobin}
%\email{webmaster@marysville-ohio.com}
%affiliation{%
%  \institution{Institute for Clarity in Documentation}
 %%\state{Ohio}
  %\country{USA}
%}

\author{Raghul Saravanan}
\affiliation{%
 \institution{George Mason University}
  \city{Fairfax, Virginia}
  \country{USA}}
\email{rsaravan@gmu.edu}

\author{Sudipta Paria}
\affiliation{%
 \institution{University of Florida}
  \city{Gainesville, Florida}
  \country{USA}}
\email{sudiptaparia@ufl.edu}

\author{Aritra Dasgupta}
\affiliation{%
 \institution{University of Florida}
  \city{Gainesville, Florida}
  \country{USA}}
\email{aritradasgupta@ufl.edu}

\author{Venkat Nitin Patnala}
\affiliation{%
 \institution{George Mason University}
  \city{Fairfax, Virginia}
  \country{USA}}
\email{vpatnala@gmu.edu}

\author{Swarup Bhunia}
\affiliation{%
 \institution{University of Florida}
  \city{Gainesville, Florida}
  \country{USA}}
\email{swarup@ece.ufl.edu}

\author{Sai Manoj P D}
\affiliation{%
 \institution{George Mason University}
  \city{Fairfax, Virginia}
  \country{USA}}
\email{spudukot@gmu.edu}

%\affiliation{%
 % \institution{Inria Paris-Rocquencourt}
 %%\country{France}
%}

%%\affiliation{%
 %\institution{Rajiv Gandhi University}
 %\city{Doimukh}
 %\state{Arunachal Pradesh}
 %\country{India}}

%\affiliation{%
 % \institution{Tsinghua University}
  %\city{Haidian Qu}
  %\state{Beijing Shi}
  %\country{China}}

%\affiliation{%
 % \institution{Palmer Research Laboratories}
 % \city{San Antonio}
 % \state{Texas}
 % \country{USA}}

%\affiliation{%
%  \institution{The Th{\o}rv{\"a}ld Group}
 % \city{Hekla}
 % \country{Iceland}}
%\email{jsmith@affiliation.org}

%\affiliation{%
%  \institution{The Kumquat Consortium}
%  \city{New York}
%\email{jpkumquat@consortium.net}


\begin{abstract}

% Org
In the evolving landscape of integrated circuit (IC) design, the increasing complexity of modern processors and intellectual property (IP) cores has introduced new challenges in ensuring design correctness and security. Recent attacks targeting hardware vulnerabilities have heightened the need for enhanced security measures throughout the design flow. The recent advancements in hardware fuzzing techniques have shown their efficacy in detecting hardware bugs and vulnerabilities at the RTL abstraction level of hardware. However, they suffer from several limitations, including an inability to address vulnerabilities introduced during synthesis and gate-level transformations. These methods often fail to detect issues arising from library adversaries, where compromised or malicious library components can introduce backdoors or unintended behaviors into the design. This gap leaves critical flaws undetected and underscores the need for more comprehensive fuzzing techniques that extend beyond the RTL level to ensure hardware integrity and security. In this paper, we present a novel hardware fuzzer, SynFuzz, designed to overcome the limitations of existing hardware fuzzing frameworks. SynFuzz focuses on fuzzing hardware at the gate-level netlist to identify synthesis bugs and vulnerabilities that arise during the transition from RTL to the gate-level. We analyze the intrinsic hardware behaviors using coverage metrics specifically tailored for the gate-level. Furthermore, SynFuzz implements differential fuzzing to uncover bugs associated with EDA libraries.
We evaluated SynFuzz on popular open-source processors and IP designs, successfully identifying 7 new synthesis bugs. Additionally, by exploiting the optimization settings of EDA tools, we performed a compromised library mapping attack (CLiMA), creating a malicious version of hardware designs that remains undetectable by traditional verification methods. We also demonstrate how SynFuzz overcomes the limitations of the industry-standard formal verification tool, Cadence Conformal, providing a more robust and comprehensive approach to hardware verification.

\vspace{1 em}

%\textcolor{red}{Be specific for the last sentence - about results.}
\end{abstract}

\maketitle

\vspace{-0.5 em}
\section{Introduction}
\label{intro}

The intricate global semiconductor supply chain demands seamless collaboration between IC designers and vendors, with various entities playing vital roles at every stage of the IC life cycle, including design, verification, fabrication, and integration \cite{icdesignflow1,icflow}. For example, the design of the Apple\textsuperscript{\tiny\textregistered} A15 chip involved 11 third-party entities to deliver sophisticated solutions \cite{apple}. However, this intricate design process is prone to trust issues, such as bugs and vulnerabilities, due to opaque interactions between entities\cite{Artenstein'17, Liu'17, Lipp'18, Kocher'18,latent}. The growing dependence on third-party intellectual property (3P-IP) blocks exponentially expands the scope for exploiting design vulnerabilities. These vulnerabilities stem from the interaction of hardware and software components in modern IC designs, especially system-on-chips (SoCs) and microprocessors \cite{dispel,paria2023divas}. The number of identified common vulnerability enumerations (CVEs) recorded in 2024 is close to 29,004, increased by 43\% compared to 2021 \cite{nistnvd}. Moreover, bugs emerging from the hardware are irreversible and pose a significant challenge to the integrity of the system.

The hardware vulnerabilities can manifest at various levels of abstraction within the design flow, including the design phase (Register Transfer Level (RTL)), the synthesis stage (gate-level netlist), and the implementation phase (physical layout) \cite{Bulck'18,Schwarz'19,Jang'16,Wojtczuk'12,Feng'22,Ghosh'23,Kang'19,Alatoun'21,Tang'22,Mohandoss'18,Vangal'08}. The sophisticated IC design flow relies on state-of-the-art Electronic Design Automation (EDA) tools \cite{Cadence, Jaspergold'23, Siemens-Questa, Cadence-Xcelium, Aldec, SymbiYosys, ABC} that facilitate seamless transitions across different hardware abstraction levels. However, the inherent nature of the hardware design flow exposes EDA tools to potential threats from adversarial entities, including malicious developers or compromised vendors. These entities, often treated as trusted, can manipulate the EDA tools to inject subtle but significant bugs, resulting in a malicious piece of hardware design \cite{mtfuzzlit}. Thus, hardware verification is essential at each abstraction in the design flow to meet the functional and security requirements \cite{mtfuzzlit,llm_survey}.

During the synthesis stage, the IP designs, expressed in Hardware Description Languages (HDL) at the RTL, are translated to an equivalent gate-level netlist representation through EDA logic synthesis tools. This process involves a variety of logic optimizations to meet performance, power, and area (PPA) requirements by mapping the HDL design to instances of standard-cell gates provided by the library vendors as binary libraries. These libraries, which are tailored to a specific technology node, represent the functional and physical properties of the gates and logic cells used in the design. The accuracy of this translation and mapping is crucial, as any inconsistencies can lead to functional errors or vulnerabilities in the final implementation phase. 

To ensure the translated design adheres to the intended behavior specified at the RTL level, formal verification methods such as Logic Equivalence Check (LEC), often leveraging techniques such as Bounded Model Checking (BMC) \cite{Azar2018SMTAN, smt}, are employed to compare the gate-level netlist with the RTL design and identify any discrepancies or unintended changes \cite{bmc}. However, these formal verification methods face significant scalability challenges and heavily rely on human expertise, limiting their efficacy in bug detection for complex designs \cite{Wang'18, Tiwari'11, Ardeshiricham'17, Li'11, Li'14, Zhang'15, Meng'22,dispel}. Hence, there is a compelling need for methodologies and tools to capture the bugs and vulnerabilities introduced during the synthesis process.

Hardware fuzzing  \cite{Laeufer'18,Li'21, Hur'21,Trippel'22,Muduli'20, Kabylkas'21, Kande'22}, inspired by software fuzzing \cite{AFL'23,microsoftrisk,ossfuzz}, has gained significant traction due to its effectiveness in detecting bugs in complex hardware designs. To date, numerous proposals have focused on developing hardware fuzzing methodologies aimed at identifying bugs at the RTL level, particularly in the designs of CPU architectures \cite{Laeufer'18,Kande'22,Canakci'23,saravanan2024emergencehardwarefuzzingcritical}. However, these methodologies are inherently limited by the chosen level of abstraction, which focuses exclusively on the pre-synthesis stage. This abstraction overlooks the transformations and optimizations that occur during synthesis, where subtle yet impactful bugs can be introduced. As a result, existing fuzzing frameworks are inadequate for detecting synthesis-induced bugs that manifest at the gate-level netlist stage. This gap underscores the necessity for innovative fuzzing approaches that can operate effectively at the gate-level netlist to verify the functional correctness between the RTL and the gate-level netlist.

\vspace{-0.5 em}

\paragraph{\textbf{Our Proposed Work:}} In this work, we propose \synfuzz, a novel hardware fuzzer that leverages fuzzing at the gate-level netlist aimed at detecting synthesis and library-related bugs. To the best of our knowledge, this is the very first work on hardware fuzzing at gate-level netlist. The input to our \synfuzz~is a gate-level abstraction of the design, where the HDL is synthesized and mapped to library cell gates provided by the standard-cell library. \synfuzz~captures intrinsic hardware behaviors associated with library cell gates, enabling the detection of subtle inconsistencies introduced during the synthesis process. 

Furthermore, we show that a maliciously modified library, designed to appear more attractive to EDA tools during the optimization and synthesis stages, can be exploited to introduce a novel threat model, termed \textbf{CLiMA (Compromised Library Mapping Attack)}. CLiMA enables flawed library mapping during the synthesis process, enabling targeted attacks on the designs, and resulting in unintended hardware implementation. We leverage CLiMA to design a malicious version of the or1200 CPU design, evading a leading formal verification tool, Cadence\textsuperscript{\tiny\textregistered} Conformal. In contrast to software fuzzers \cite{libfuzz,AFL'23,ossfuzz} that rely on crashes, and hardware fuzzers that depend on golden models \cite{Laeufer'18,hyperprop} or ISA for validation \cite{Canakci'23,Kande'22}, EDA libraries lack any equivalent golden models. In this work, we address this challenge by identifying library bugs in the context of differential fuzzing. As such, \synfuzz~\textit{1) supports conventional hardware and design verification flow 2) detects synthesis and library vulnerabilities during the synthesis stage 3) does not require extensive design or EDA tool knowledge 4) is scalable to large and complex designs.} In summary, our key contributions are :

\vspace{-0.1 em}

\begin{itemize}
\vspace{-0.1 em}
    \item We propose our new hardware fuzzer, \synfuzz~that leverages state-of-the-art gate-level fuzzing at post-synthesis level to unveil \textcolor{black}{RTL}, synthesis and library bugs(Section \ref{proposed}).
    \item  This work introduces a new class of synthesis attack model CLiMA (Section \ref{clima}), leveraging malicious library mapping through the logic synthesis tools.
     \item We leverage CLiMA to inject bugs in the or1200 CPU design that evades formal verification methods such as Logic Equivalence Check (Section \ref{random}, \ref{targetclima}). 
    \item We extensively evaluate our fuzzer, \synfuzz, on four popular and complex open-source CPU designs: 1) the or1200 processor (OpenRISC ISA) \cite{RISC}, 2) the IBEX processor (RISC-V ISA) \cite{RISC-V}, 3) the PicoRV32 processor (RISC-V ISA) \cite{RISC-V}, and 4) the MIPS processor \cite{mips}. Additionally, we test it on cryptographic cores such as RSA from TrustHub \cite{6657085} and Data Encryption Standard (DES) \cite{cep}, a Digital Signal Processor (DSP) \cite{CADIEEE}, and several IP peripherals, including UART \cite{CADIEEE}, GPIO \cite{CADIEEE}.
    \item \synfuzz~found 7 new synthesis bugs across three processors and one cryptographic core, with all 7 being newly discovered bugs (Section \ref{bugs}). In addition, \synfuzz~ found 10 existing synthesis translation bugs in popular open-source EDA synthesis tool Yosys. To evaluate the efficacy of our fuzzer, we utilized a leading formal verification tool, Cadence's Conformal Logic Equivalence Check \cite{LEC_cadence}. \synfuzz~addresses the limitations of the Conformal tool, including scalability issues and its heavy reliance on human expertise (Section \ref{leceval}).

\end{itemize}

\vspace{-1 em}

\vspace{-0.5 em}
\section{Background}
\label{background}
In this section, we furnish the necessary background on hardware fuzzing, and the hardware development cycle.

\begin{figure*}[h]
% \vspace{-1em}
  \centering
  \includegraphics[width=0.8\linewidth]{
  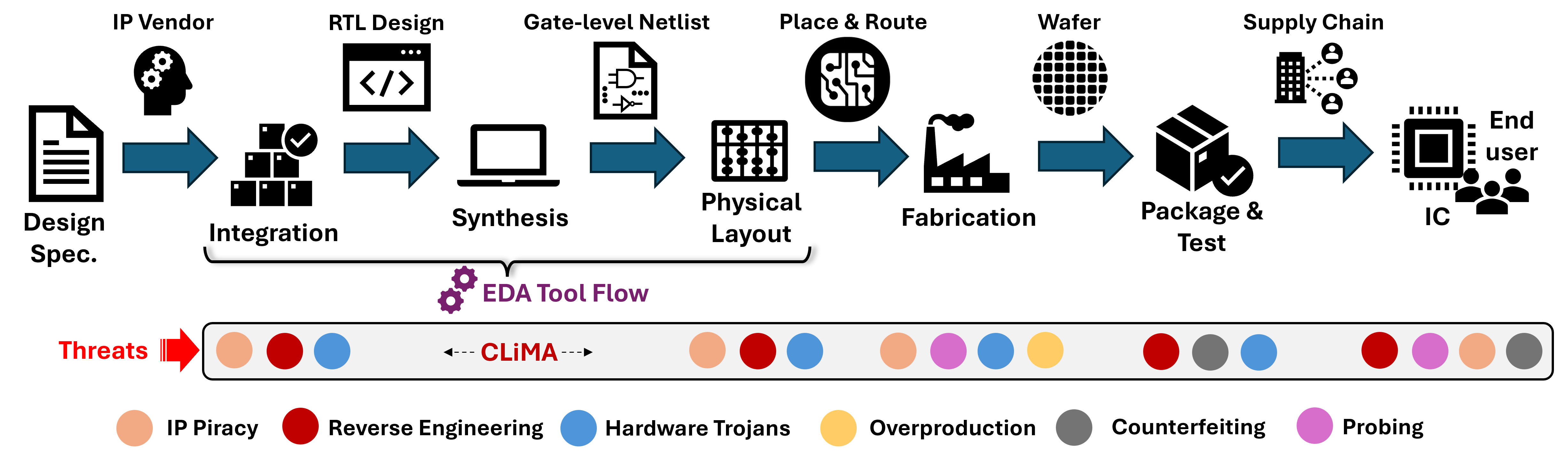
  } % \vspace{-1em}
  \caption{Threat Space in the Modern IC Design Flow}
  \label{fig:overview}
%  \vspace{-0.5em}
\end{figure*}

\vspace{-0.5 em}

\subsection{From Design to Threats: Unpacking the Modern IC Lifecycle} \label{hwdevpipeline}

The recent advancements in technology and the rising demand for enhanced performance, energy efficiency, and robust security have significantly increased the complexity of the modern IC design process. However, throughout the IC development lifecycle, various threats can emerge at different stages, from design to deployment, as shown in Figure \ref{fig:overview}. Starting with defining system-level requirements and specifications in the design stage, the functional, performance, and security objectives are outlined for the IC. IP vendors generate the register-transfer level (RTL) code using hardware description languages such as Verilog, VHDL, and SystemVerilog, which align with the design descriptions through data flow between registers and the logic operations that occur during each clock cycle. The RTL design undergoes thorough verification to ensure it meets the functional requirements \cite{watermarking,Sasan21,mtswarup,Rajendran2013SecurityAO}. This verification is typically carried out using RTL simulations via EDA tools like Cadence Xcelium, Synopsys VCS \cite{Cadence-Xcelium,Synopsys,Cadence, Synopsys-VCS} which apply test vectors to check for correctness and identify any functional bugs in the design. During this stage, risks such as inaccurate requirements, design flaws, and Trojan insertion can result in IP theft, data leakage, and backdoor exploitations and several countermeasures are proposed to protect against diverse attacks \cite{Bulck'18,Schwarz'19,Jang'16,Wojtczuk'12,Feng'22,harpoon,pip_2023,Ghosh'23,spell,Kang'19,paper1,hipr,Alatoun'21,Tang'22,Mohandoss'18,Vangal'08}.

In the next phase, EDA tools perform synthesis that translates the RTL design into a gate-level representation. This step maps the design onto fundamental hardware components such as logic gates, flip-flops, buffers, and other standard cells from the technology library provided by the semiconductor foundry. However, it is essential to ensure that the gate-level design meets both functional and timing requirements \cite{Farzana'19} requiring through verification efforts. The current state-of-the-art techniques, including formal methods, functional simulations, and fault simulations, are employed to compare the design against the RTL specifications \cite{Cadence,Jaspergold'23,Rajendran2013SecurityAO}. During synthesis and post-synthesis verification, issues such as timing violations, incomplete verification, and malicious modifications can arise \cite{cascade}. The adversary can exploit the EDA tool optimizations during synthesis to introduce bugs that propagate to later stages, resulting in IP piracy, denial-of-service, confidentiality violations, etc \cite{Dunbar2014DesigningTE}. \textit{Based on these threat spaces identified during the synthesis stage, this serves as the primary motivation to extend fuzzing methodologies to the gate-level netlist}. By targeting this critical abstraction level, it becomes possible to identify synthesis-induced bugs and vulnerabilities that are otherwise missed by traditional RTL-level fuzzing approaches.

%and also suffers from limited observability in terms of debugging and monitoring internal signals. 

The design then moves into the physical design phase, where the gate-level netlist is converted into a layout, followed by physical verification, preparing the design for tape-out and subsequent manufacturing at the foundry. At this stage, all verification occurs in a pre-silicon environment, meaning the design is represented as RTL, gate-level, or layout models, not yet as physical silicon. A key advantage of pre-silicon verification is the high degree of observability, which allows engineers to monitor and debug the design thoroughly. This level of visibility is not possible in post-silicon validation. The physical design and fabrication stages are vulnerable to threats such as overproduction, counterfeit ICs, hardware Trojans, and probing attacks, all of which can significantly undermine the chip's reliability and trustworthiness \cite{book1}.

Once the chip is fabricated, post-silicon verification is employed which is more complex and costly than pre-silicon verification, offering limited observability. Its primary goal is to confirm that the manufactured silicon meets the design specifications and functions as intended. Throughout both the pre-and post-silicon verification stages, each level of the design is compared to its previous respective Golden Reference Model (GRM) to ensure accuracy and reliability.

\subsection{Fuzzing} \label{hwfuzz}

\begin{figure}[htb!]
\vspace{-1em}
  \centering
  \includegraphics[width=1\linewidth]{
  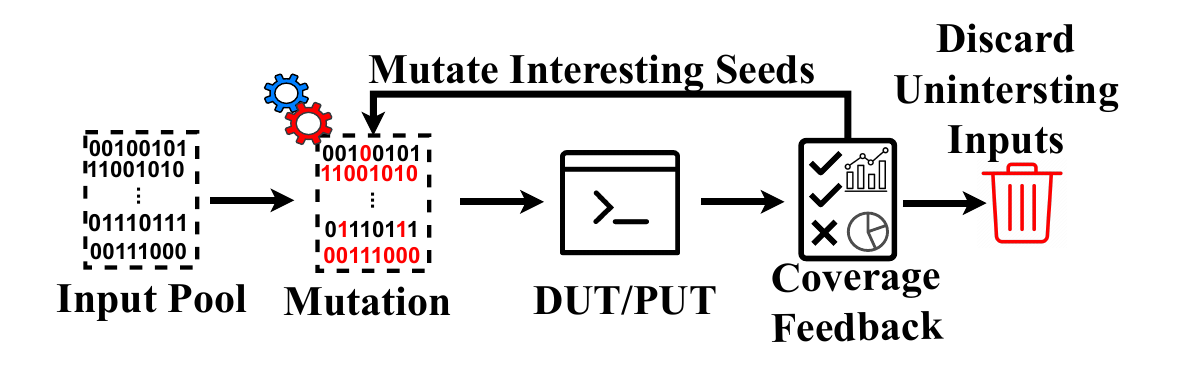} \vspace{-1em}
  \caption{Overview of Fuzzing
  %\textcolor{red}{You have white space in fig, you need to remove it.}
  % \textcolor{blue}{acknowledged}
  % \cite{9} \textcolor{red}{Did you redraw this figure?\textcolor{violet}{Done}}
  } \label{fig:fuzzfig}
  \vspace{-0.5em}
\end{figure}

Fuzzing involves bombarding randomized input seeds to the Program Under Test (PUT)/Design Under Test (DUT), intending to trigger the existing bugs and vulnerabilities. The generated input seeds are further mutated based on the coverage feedback obtained from the DUT/PUT under investigation, as shown in Figure \ref{fig:fuzzfig}. The popular mutation algorithms used by fuzzers are  American Fuzzy Lop (AFL) mutation algorithms \cite{AFL'23} such as bit-flip, swap, arithmetic, etc. The coverage feedback aids in steering the fuzzer to generate effective seeds for improved coverage and discard uninteresting inputs. Based on the various mechanisms to obtain coverage feedback, fuzzers are classified as white-box, grey-box, and black-box fuzzing. The outcomes from the DUT/PUT are analyzed for any crash or GRM implementation to detect the bugs. Fuzzing has proven its efficiency in finding vulnerabilities in software \cite{microsoftrisk,ossfuzz,libfuzz,Libfuzzer'21} and hardware platforms \cite{Cadence, Wile'05,Clarke'12, Hicks'15, Guo17,Guo2017,Guo3017,Farahmandi,   bmc, Drzevitzky2010,Z3}. Industry-based fuzzing frameworks such as Google's OSS-Fuzz \cite{Serebryany'17} and Microsoft's Security Risk Detection \cite{microsoftrisk} have proved their efficacy and effectively identified a plethora of security vulnerabilities. The popular software fuzzer, AFL \cite{AFL'23}, is predominantly used in software fuzzing and is predominantly used in the majority of the software fuzzing techniques \cite{microsoftrisk,ossfuzz,libfuzz,Libfuzzer'21}.

In the context of hardware fuzzing, the fuzzing target is at different levels of abstraction including RTL or gate-level netlist or physical layout. There are three ways to fuzz the hardware: 1) fuzzing hardware like software \cite{Trippel'22,Verilator}  2) direct fuzzing on hardware \cite{Laeufer'18} 3) fuzzing hardware like hardware \cite{Kande'22,hybrid,borkar2024whisperfuzz}. The existing hardware fuzzers primarily focus on the RTL as their chosen level of abstraction to identify bugs.

\begin{formal}
    \textbf{Observation O1:} The chosen level of hardware abstraction and fuzzing methodologies by existing hardware fuzzers are incapable of discovering synthesis bugs, and EDA library vulnerability exploits. 
\end{formal}

\begin{figure}[htb!]
\vspace{-1em}
  \centering
  \includegraphics[width=1\linewidth]{
  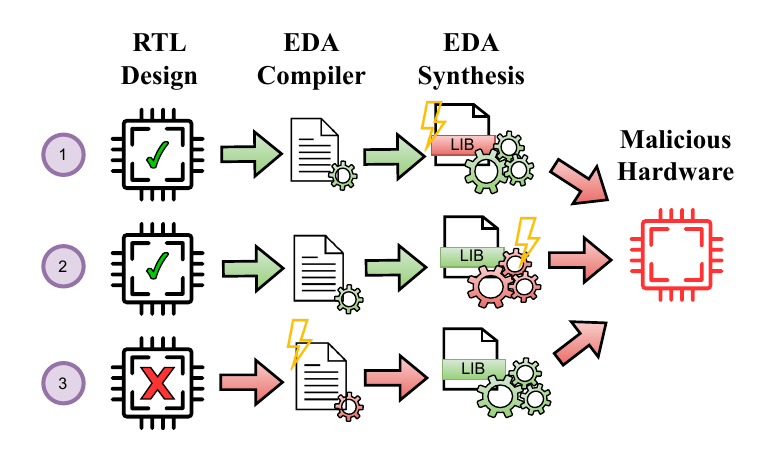} \vspace{-1em}
  \caption{Overview of our Threat Models. 1) Malicious Library Mapping Exploitation 2) EDA Synthesis Tool Exploitation 3) EDA Compiler Exploitation
  %\textcolor{red}{You have white space in fig, you need to remove it.}
  % \textcolor{blue}{acknowledged}
  % \cite{9} \textcolor{red}{Did you redraw this figure?\textcolor{violet}{Done}}
  } \label{fig:threatmodels}
  % \vspace{-1.5em}
\end{figure}

% \vspace{-1 em}

 \section{The Genesis of Synthesis Bugs: Foundations for Building Effective Fuzzers} \label{genesis}

The primary challenge in constructing an effective fuzzer for detecting synthesis bugs lies in establishing a comprehensive understanding of the threat model and addressing the inherent complexities of fuzzing to ensure efficient bug detection. For this purpose, in this section, we illustrate our threat model in Figure \ref{fig:threatmodels}, followed by the prevailing challenges to create a hardware fuzzer to detect the bugs in Section \ref{Challenges}.

% \vspace{-1 em}

\subsection{Threat Model}

\paragraph{\textbf{Our Threat Model Assumptions:}} We assume a scenario where a malicious hardware developer exploits EDA tools or a compromised EDA library vendor leverages the optimization configurations of synthesis tools to craft a malicious hardware design.

\paragraph{\textbf{Malicious Library Mapping:}} A legitimate RTL design is provided; however, during the EDA synthesis stage, a compromised library binary is provided to foster compromised logic mapping leading to malicious hardware as shown in Figure \ref{fig:threatmodels} \circled{1}. The EDA compiler and synthesis process appear legitimate but are tainted by the corrupted library.

\paragraph{\textbf{EDA Synthesis Tool Exploitation:}} The RTL design is legitimate and free from malicious intent. However, a malicious hardware developer exploits the EDA tools during the synthesis stage to inject bugs as in Figure \ref{fig:threatmodels} \circled{2}. This malicious intent involves manipulating the synthesis process to introduce vulnerabilities such as deleting modules or instances, creating multiple drivers for a single signal, or driving certain pins with fixed logic constants. These actions result in unintended design behavior, leading to functional errors in the gate-level netlist. The trust placed in the EDA tool \cite{SymbiYosys,Cadence,Synopsys}, often considered a trusted entity in the design flow, enables malicious activity by providing an avenue for hardware developers to exploit the synthesis process.

% \textcolor{red}{The difference between the above two are not clear. } \textcolor{blue}{Done!}

\paragraph{\textbf{EDA Compiler Exploitation:}} In this threat model, an attacker crafts specific HDL behaviors that remain undetected during RTL compilation but manifest in the synthesized design. These issues are intentionally subtle, allowing them to bypass checks during the EDA Compiler stage, which treats the RTL as legitimate, as shown in Figure \ref{fig:threatmodels} \circled{3}. Such flaws only surface during synthesis when the design is transformed into its gate-level representation, potentially leading to functional discrepancies or security vulnerabilities.

% \textcolor{red}{From looking at Fig 3, it looks like a bug at RTL level.}

% \vspace{-1 em}
\subsection{Prevailing Challenges} \label{Challenges}
\begin{formal1}
\textbf{Challenge C1:} Determine suitable hardware abstractions and input representations 
\end{formal1}

To detect the bugs associated with the above threat models, determining the level of hardware abstraction provided to the fuzzer is critical. The fuzzing target is typically a model of the hardware design represented at the architecture level, RTL level, post-synthesis netlist level, and physical level. In addition, the hardware fuzzer should generate appropriate inputs in the format corresponding with the targeted hardware abstraction level. Thus, for efficient bug and vulnerability detection, the key challenges lie in \textit{determining the appropriate hardware abstraction level for fuzzing and crafting input representations} that optimize the discovery of vulnerabilities within the design flow.

\begin{formal1}
\textbf{Challenge C2:} Extract suitable coverage metrics 
\end{formal1}

The second challenge is to extract suitable coverage feedback mechanisms with the selected hardware abstraction level that can detect bugs and vulnerabilities. The chosen coverage metrics should capture complex hardware behaviors at the chosen abstraction level. This helps the fuzzers to cover critical regions of the hardware design and also assists in generating efficient inputs. Hence, \textit{the extraction of coverage metrics for feedback mechanism is crucial. }

\begin{formal1}
\textbf{Challenge C3:} To detect bugs through a reliable reference model
\end{formal1}

The third challenge pertains to the detection of bugs, which relies upon the chosen reference model to uncover discrepancies between the expected behavior and the actual design, highlighting bugs and vulnerabilities that arise during the synthesis process. Hence, \textit{selecting an accurate and reliable reference model is crucial for effectively identifying bugs. }

\begin{formal1}
\textbf{Challenge C4:} Exploit the bugs and the vulnerabilities introduced through EDA tools
\end{formal1}

The final and critical challenge is to exploit the bugs and vulnerabilities or malicious modifications introduced at the synthesis stage of the hardware design flow that can evade traditional verification methods such as LEC, which are commonly used in conventional industry practices.

\section{Design of SynFuzz}
\label{proposed}

\begin{figure}[htb!]
	\vspace{-1em}
	\centering
	\includegraphics[width=1\linewidth]{
		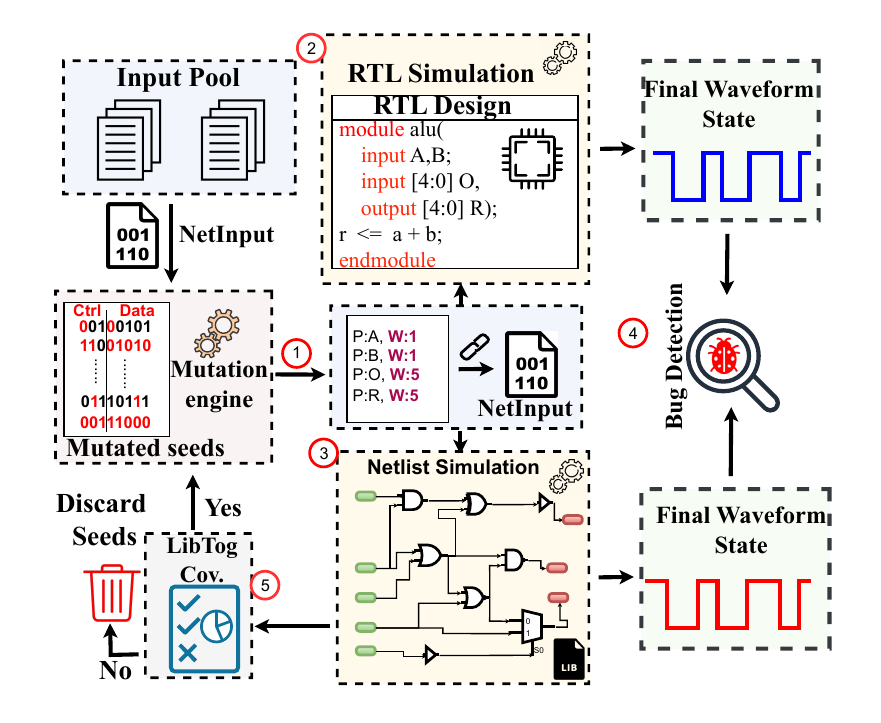} \vspace{-1em}
	\caption{Overview of SynFuzz}

	 \label{fig:SynFuzz2}
	\vspace{-1em}
\end{figure}

To address the challenges outlined in Section \ref{Challenges} and detect bugs associated with the above threat models in Section \ref{genesis}, we present a novel hardware fuzzer \synfuzz~  as shown in Figure \ref{fig:SynFuzz2}. We fuzz the hardware at the gate-level netlist abstraction, which is mapped to library cells by the EDA synthesis tool, representing the equivalent form of the RTL design, to detect synthesis bugs. The \synfuzz~ framework is adaptable to the conventional hardware design flow. We first introduce how \synfuzz~ performs mutations and generates an input seed format, \textit{NetInput}, which is compatible with both the RTL and the gate-level netlist (Section \ref{seedgen}). Next, we illustrate how \synfuzz~ compiles and simulates the RTL and gate-level netlist using these inputs (Section \ref{sim}). Moving forward, we propose a coverage metric, which captures intrinsic hardware behaviors at the gate-level netlist to refine and guide the mutation of interesting seeds (Section \ref{libtog}). Lastly, we describe how \synfuzz~ cross-verifies the execution results from the RTL and gate-level netlist to identify bugs effectively (Section \ref{bugdetection}).

\paragraph{\textbf{Overview of Workflow:}} The overall framework of SynFuzz is shown in Figure \ref{fig:SynFuzz2}. First, the randomly generated initial seeds are mutated (\circled{1}), and SynFuzz runs both RTL and gate-level netlist simulation using the input (\circled{2}, \circled{3}). After the completion of the simulations, SynFuzz compares the final traces of the execution results of these simulations to detect bugs (\circled{4}). The seeds are further mutated using the coverage metrics to increase the exploration of uncovered hardware behaviors and improve the detection of synthesis bugs (\circled{5}). 

% \vspace{-1 em}

\subsection{NetInput Seed Generation} \label{seedgen}
To address Challenge C1 in Section \ref{Challenges} regarding input representation, \synfuzz~ proposes an innovative input format, \textit{NetInput}. The gate-level netlist, which represents the synthesized equivalent of the RTL design, retains the same input ports (P) and width (W), which are concatenated to form a series of bits as shown in Figure \ref{fig:SynFuzz2}. % \textcolor{red}{Provide a better example. I mean try to have a separate figure for this.}
This input format is specifically designed to provide identical inputs to both the RTL and the gate-level netlist abstraction, ensuring consistency and compatibility across these two levels of hardware representation. NetInput ensures that any discrepancies or bugs identified during the fuzzing process can be directly attributed to the synthesis bugs.  
NetInput includes all input space that the RTL and gate-level netlist can process during the fuzzing process. For a given NetInput, SynFuzz incorporates AFL-style mutation techniques \cite{AFL'23}. For an efficient mutation process, the user annotates the control and data bits to the mutation engine. As shown in Figure \ref{fig:SynFuzz2} \circled{1}, two distinct mutation engines are employed: one tailored for control bits and the other for data bits. Mutating the control bits helps uncover different control paths while mutating the data bits focuses on exploring the datapath.

% \textcolor{red}{How do you know/determine which bits belong to control and data paths? Also, is the input generation random or do you have any logic behind it? } \textcolor{blue}{done !}

\subsection{RTL and Netlist Simulation} \label{sim}

\synfuzz~ leverages state-of-the-art EDA simulation tools to perform simulations at both the RTL and gate-level netlist abstractions. The mutated bit-vectors, which serve as input stimuli, are loaded onto the testbenches to drive simulations of the hardware designs, as depicted in Figure \ref{fig:SynFuzz2} (\circled{2}, \circled{3}). 
For the gate-level netlist simulation, \synfuzz~ incorporates the library environment alongside the netlist. This environment is crucial for resolving and inferring the mapped instances of library cell gates, which are provided by the library vendor. These mapped instances reflect the physical implementation of the design and play a vital role in synthesis-related bugs introduced during the mapping process. \synfuzz~ focuses primarily on the intrinsic hardware behaviors at the gate-level netlist during the fuzzing process. In the following section, \synfuzz~ introduces a coverage metric designed to capture the unique characteristics of the hardware at the gate level.

%SynFuzz leverages state-of-the-art EDA simulation tools for the RTL and gate-level netlist simulations. The mutated bit-vectors are loaded onto the testbenches to simulate the hardware abstractions as shown in Figure \ref{fig:SynFuzz2} (\circled{2}, \circled{3}). For the gate-level netlist simulation, the library environment is provided alongside the netlist to infer the mapped instances of library cell gates provided by the library vendor. SynFuzz focuses on the intrinsic hardware behaviors at the gate-level netlist during the fuzzing process. In the following section, we propose a coverage metric designed to capture these characteristics.

\subsection{Library Toggle Coverage} \label{libtog}

The hardware design represented at the RTL encompasses diverse and rich combinational and sequential circuits. The combinational circuits perform logic operations where the output depends solely on the current inputs, such as adders, multiplexers, and decoders, while sequential circuits, on the other hand, incorporate memory elements, where the output depends on both the current inputs and the stored state, as seen in flip-flops, registers, and finite state machines. During the synthesis process, these combinational and sequential components are mapped to their equivalent library gate cells provided by the standard-cell library. In the following, we demonstrate the efficacy of our proposed coverage metric in capturing the hardware characteristics associated with the mapped library gate cells.\\

\textbf{Case Study:} We illustrate with a case study using a code snippet presented in Listing 1, similar to D-link hardware authentication \cite{CVE'21}. Firstly, we begin by describing the expected functionality and then outline the bugs. Finally, we detail how the Library Toggle coverage metrics are utilized to detect these bugs.

For Listing 1, the system should grant access by verifying the password (\verb|pwd|) against the stored value in memory (\verb|mem_123|), even when the superuser mode is enabled. This ensures that access is granted only to authorized users and prevents unauthorized access to the system. In addition, when the debug mode is enabled (\verb|debug|), the system should check whether the input data (\verb|in_data|) matches with the input data stored in the memory (\verb|mem_456|). This access check prevents unauthorized or invalid inputs from being accepted under the guise of debug mode. \\

\begin{lstlisting}[caption=Access control code snippet,frame=single,
	,basicstyle=\ttfamily\footnotesize,keywordstyle=\color{red},language=VHDL,breaklines=true,showstringspaces=true,keepspaces=false]
	process (sudo_en, pwd, mem_123, access_level, debug, in_data, mem_456)
	begin
	if sudo_en = '1' then
	access_granted <= '1';  
	elsif pwd = mem_123 then
	access_granted <= '1'; 
	else
	access_granted <= '0'; 
	end if;
	
	if debug = '1' then
	input_valid <= '1';  
	elsif in_data = mem_456 then
	input_valid <= '1'; 
	else
	input_valid <= '0';  
	end if;
	if access_granted = '1' or input_valid = '1' then
	system_access <= '1';  
	else
	system_access <= '0'; 
	end if;
	end process;
\end{lstlisting} \label{snippet}  

\vspace{0.1in}

\begin{figure}[htb!]
	\vspace{-1em}
	\centering
	\includegraphics[width=1\linewidth]{
		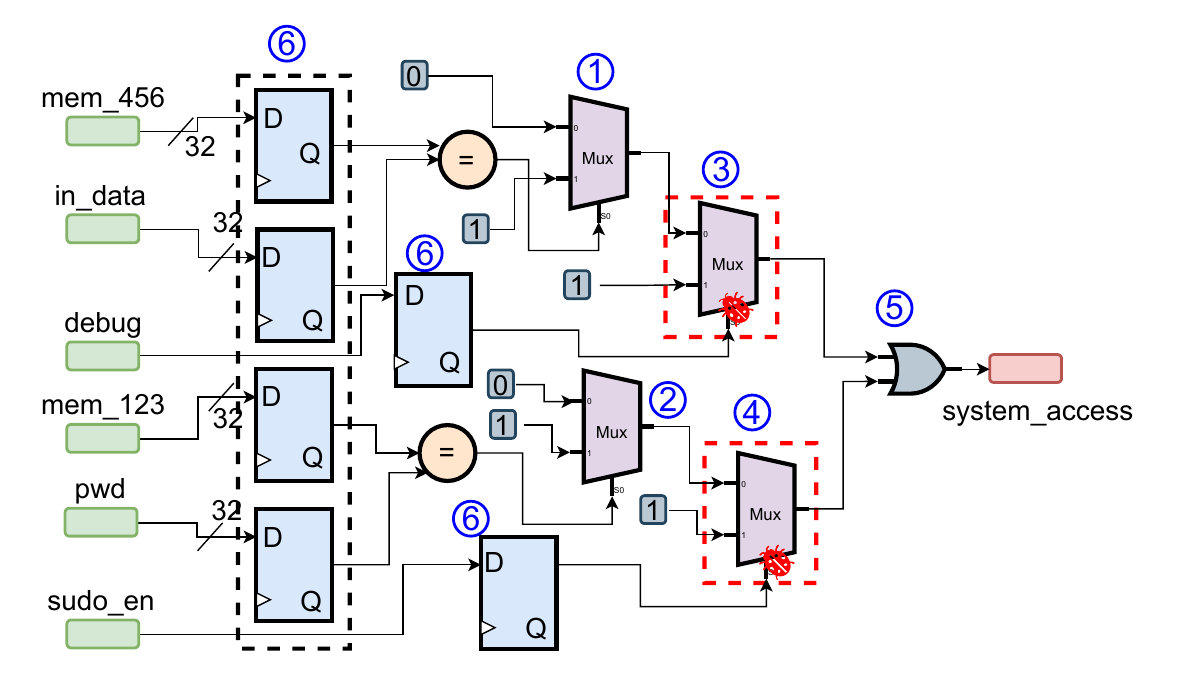} \vspace{-1em}
	\caption{Hardware Design for Listing 1}
		
	 \label{fig:libtog}
	\vspace{-0.5em}
\end{figure}

The EDA tools synthesize this RTL code into equivalent library gate cells, as illustrated in Figure \ref{fig:libtog}. The multiplexers \circled{1} and \circled{2} handle the matching operations for the password \verb|pwd| and input data \verb|in_data| to determine access eligibility. Multiplexers \circled{3} and \circled{4} evaluate the selection of superuser mode or debug mode. The combinational logic in  \circled{5} controls the final system access decision, while \circled{6} serves as a register to hold the input values.

For the design in Figure \ref{fig:libtog}, there exist two bugs, \textit{b1} and \textit{b2}, similar to D-Link hardware authentication. These are analogous to CVE-2021-34696 \cite{CVE'21} (access control bypass), CVE-2021-1071 \cite{CVE'21b} (access control bypass), and CVE-2021-1088 \cite{CVE'21a} (user utilizing debug mode with insufficient access control). For the bug \textit{b1}, as observed in Line 4 of the code snippet, if the \verb|sudo_en| flag is enabled, regardless of the entered password (\verb|pwd|), the password check flag (\verb|pwd_check|) is set, effectively bypassing the password validation. This compromises the security of the system by allowing unauthorized access. The source of this bug is the multiplexer \circled{3}. Similarly, for the bug \textit{b2} if the \verb|debug| flag is set, then the input check (\verb|in_check|) is bypassed which is associated with \circled{4}.

In \circled{1}, \circled{2}, \circled{3}, \circled{4} all the inputs to the MUX and the corresponding select signals should be monitored for their correctness. For the combinational logic, \circled{5}, it is essential to verify the intermediate computations and ensure that the logic correctly propagates the signals to grant system access. The registers \circled{6} can take all possible combinations for the system access check. 

For all of these scenarios, we utilize the \textbf{library toggle coverage} of the gate cells to monitor and analyze the toggling behavior of the library gate cells. This metric indicates whether a particular instance of a mapped library cell is activated or toggled during the simulation process. By tracking the toggle activity of the instances \circled{3}, \circled{4}, \synfuzz~ can detect the bugs \textit{b1} and \textit{b2} in the Listing 1. Furthermore, for the multiplexers \circled{1} and \circled{2}, the input data consists of constant values, which are declared using \verb |assign| statements, such as \verb|assign c1 = 1'b0|. These expression values play an important role in granting access when there is a password match, and the input data is valid. We use \textbf{expression coverage} to verify whether the values assigned to \circled{1} and \circled{2} are legit. \synfuzz~ leverages state-of-the-art industry standard EDA tools such as Synopsys VCS \cite{Synopsys,Synopsys-VCS} or Cadence Xcelium \cite{Cadence-Xcelium} to extract the library toggle coverage and expression coverage metrics. These tools have been used in the industry and academia over the decades which aligns with the traditional design verification flows. Thus, \textit{SynFuzz's} coverage metrics aid in detecting the bugs \textit{b1} and \textit{b2} through the library toggle ane expression coverage metric.

% \vspace{-1 em}

\subsection{Bug Detection} \label{bugdetection}

Upon the completion of executing both the RTL and gate-level netlist for a given \textit{NetInput}, \synfuzz~ initiates the verification phase. Similar to traditional hardware verification, where the gate-level netlist is compared against the RTL, \synfuzz~ compares the waveform traces of the RTL and gate-level netlist to detect any discrepancies as shown in Figure \ref{fig:SynFuzz2} \circled{4}. Initially, \synfuzz~ checks whether the output register waveforms of the RTL and gate-level netlist align. If a mismatch is detected, \synfuzz~ flags the \textit{NetInput} for the presence of a bug, which is then manually analyzed to identify its root cause. This process ensures that any inconsistencies introduced during synthesis are accurately identified and recorded for further analysis.

\begin{formal1}
	\textbf{Challenge C5:} Absence of GRM for EDA Libraries
\end{formal1}

However, note that \synfuzz~ relies on library cells during the synthesis process to identify bugs, which implies two limitations \textit{1) Absence of GRM for EDA Library 2) Variability in Mapping Algorithms of the EDA tool and Library Characteristics.} Although the gate-level netlist is mapped to library cells to represent the equivalent RTL design, different EDA tools employ distinct algorithms for mapping the design, and library cells often vary in their logical and physical characteristics (e.g., timing, power, area). Unlike software fuzzers that rely on crashes to sanitizers and hardware fuzzers that rely on golden models, there is no GRM for EDA libraries, which introduces significant challenges in validating the instances of library cells.

\begin{formal1}
	\textbf{Challenge C6}: Variability in Mapping Algorithms and Library Characteristics
\end{formal1}

To address these challenges, in conjunction with \synfuzz~, we introduce \textit{DiffLib}, which extends the fuzzing process by performing differential analysis across various libraries and tools. Despite this extension, the outputs of the gate-level netlist are still compared solely with the RTL trace. By addressing these limitations, \textit{DiffLib} complements \synfuzz~, providing a holistic approach to detecting bugs in the synthesis process, especially those arising from tool or library-specific inconsistencies.

\begin{figure}[htb!]
	\vspace{-1em}
	\centering
	\includegraphics[width=1\linewidth]{
		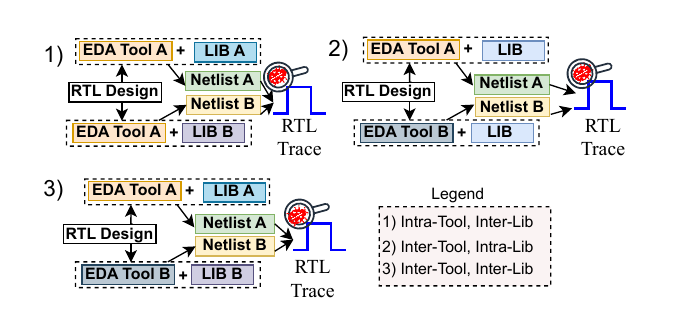} \vspace{-1em}
	\caption{Configurations of DiffLib. 1) Intra-Tool, Inter-Lib 2) Inter-Tool, Intra-Lib 2) Inter-Tool, Inter-Lib}
		
		% \textcolor{blue}{acknowledged}
		% \cite{9} \textcolor{red}{Did you redraw this figure?\textcolor{violet}{Done}}
	 \label{fig:difflib}
	\vspace{-0.1em}
\end{figure}

\vspace{-1 em}

\subsection{DiffLib} \label{difflib}

We find that differential fuzzing can be applied in three different ways \textit{1) Intra-Tool, Inter-Lib 2) Inter-Tool, Intra-Lib 3) Inter-Tool, Inter-Lib} as shown in Figure \ref{fig:difflib}. These approaches are designed to systematically identify inconsistencies in netlist generation by varying the tools and libraries used during the synthesis process.

\paragraph{\textbf{Intra-Tool, Inter-Library:}} In this approach, netlists are generated using the same EDA synthesis tool with two different standard cell libraries, as shown in Figure \ref{fig:difflib} \circled{1}. The goal is to identify inconsistencies caused by library-specific variations, such as differences in logical functions, timing, power, or area. By keeping the synthesis tool constant, the impact of these variations on the final netlist can be isolated. For instance, two libraries may implement the same logic gate with subtle structural differences, leading to discrepancies in the synthesized netlist. This method provides insight into the synthesis tool's ability to handle diverse library properties while ensuring consistent functionality.

\paragraph{\textbf{Inter-Tool, Intra-Library:}} 

This approach focuses on using different synthesis tools while keeping the standard cell library constant, as shown in Figure \ref{fig:difflib} \circled{2}. The aim is to detect discrepancies introduced by differences in how each synthesis tool interprets and handles the same library. Since EDA tools are developed by different vendors, they may employ unique optimization strategies, heuristics, and algorithms, leading to variances in the generated netlists even when using identical library files. This method highlights tool-specific behaviors and helps identify potential issues that may arise from discrepancies in synthesis outcomes when transitioning between tools in a design workflow.

\paragraph{\textbf{Inter-Tool, Inter-Library:}}  The Inter-Tool, Inter-Library approach represents the most comprehensive method, combining the variability of both synthesis tools and libraries. By testing with different EDA tools and libraries, this approach uncovers compounded discrepancies that arise from the interaction between tool-specific behaviors and library-specific characteristics. For instance, one tool might prioritize optimizations on power, and the libraries may offer differing trade-offs for these metrics. This combined analysis provides a holistic view of the variations and inconsistencies that can occur in practical design flows, offering insights into the compounded impact of tool and library diversity.

\section{Evaluation}
\label{eval}

\begin{table*}[h!]

    \centering
    \caption{ {Bugs detected in the IP designs by \synfuzz } }
    \vspace{-1em}
    \label{tb1:bugreport}
    \scalebox{0.8}{
    
    \begin{tabular}{|c|l|l|c|c|}
    \hline 

   \textbf{Design} & \textbf{Bug Description} & \textbf{Location} &  \textbf{Relevant CWE(s)} &  \textbf{New Bug ?}  \\ \hline
    
    \textbf{RSA \cite{6657085}} & \textbf{B1}: Multiple sources driving the modular arithmetic operations & Frontend & CWE-665, CWE-1419 & \cmark  \\

    \hline

    \textbf{PicoRV32 \cite{RISC-V}}  & \textbf{B2}: Incorrect memory calculation operations in Memory Interface &  Memory Interface & CWE-131, CWE-119 & \cmark \\ 
    & \textbf{B3}: Incorrect data forwarding to co-processors & Co-Processor Interface & CWE-1422 & \cmark\\
    \hline

    \textbf{MIPS \cite{mips}} & \textbf{B4}: Incorrect implementation of sum operations & ALU & CWE-682, CWE-190 & \cmark \\

    \hline
    
    \textbf{or1200 \cite{RISC}}  & \textbf{B5}: Incorrect memory data retrieval from LSU  & Frontend & CWE-125, CWE-119 & \cmark \\
    & \textbf{B6}: Failure of PC updation & Frontend  & CWE-451  & \cmark  \\
    & \textbf{B7}: Inaccurate PC updation when NPC vales in SPR changes  & Program Counter  & CWE-221, CWE-664 & \cmark  \\
   
    \hline

    \end{tabular}
    }
\end{table*}

The implementation of our proposed fuzzer, \synfuzz, seamlessly integrates with the conventional industry-standard IC design and verification flow, enabling efficient bug detection. The experiments were conducted on a 48-core Intel Xeon processor with 512 GB of RAM running RHEL Linux Operating System (OS). 
We extensively evaluate our fuzzer, \synfuzz, on four popular and complex open-source CPU designs: 1) the or1200 processor (OpenRISC ISA) \cite{RISC}, 2) the IBEX processor (RISC-V ISA) \cite{RISC-V}, 3) the PicoRV32 processor (RISC-V ISA) \cite{RISC-V}, and 4) the MIPS processor \cite{mips}. Additionally, we test it on cryptographic cores such as RSA from TrustHub \cite{6657085} and Data Encryption Standard (DES) \cite{cep}, a Digital Signal Processor (DSP) \cite{CADIEEE}, and several IP peripherals, including UART \cite{CADIEEE}, GPIO \cite{CADIEEE}. The or1200 processor is an OpenRISC based 32-bit, 5-stage pipeline processor used in academic research for multiple decades. IBEX is a 32-bit, 2-stage pipeline processor based on the RISC-V architecture. PicoRV32 is a compact 32-bit RISC-V processor designed for resource-constrained applications. The MIPS processor, a widely studied processor in academia, serves as a popular reference for educational and research purposes.

% \vspace{-1 em}
\subsection{Experimental Setup} \label{setup}

The proposed components of \synfuzz~are designed using Python scripts and custom TCL scripts to integrate the components.

\paragraph{\textbf{RTL and Netlist Simulation}:} We simulate the target hardware using a leading industry standard EDA tools, Synopsys VCS \cite{Synopsys-VCS} and Cadence Xcelium \cite{Cadence-Xcelium}. These tools support a variety of hardware abstraction levels including RTL, gate-level, and transistor-level. To extract and analyze the coverage metrics, we deployed custom TCL and Python scripts. For deriving the netlist, we utilized state-of-the-art commercial industry standard EDA synthesis tools, Synopsys Design Compiler \cite{Synopsys} and Cadence Genus \cite{Cadence}, which efficiently translate the given RTL design into a gate-level representation with high performance. In addition, we also used a popular open-source EDA tool Yosys \cite{SymbiYosys}, which is widely adopted in the OpenROAD \cite{open} flow for RTL synthesis and netlist generation. The above mentioned EDA tools are widely used in industry and academia for hardware research and development. They are capable of synthesizing large-scale designs and can seamlessly scale to complex CPU cores and full SoC architectures, making them suitable for evaluating real-world, high-complexity hardware systems.  

These tools were provided with two open source libraries 45nm \cite{open} and skywater130nm \cite{open} while we use it for DiffLib. Note that, we simulate the netlist with Zero Delay Simulation (ZDS) condition. ZDS simulates the netlist without annotating any timing data. It is mainly meant for checking and validating the functionality of the design once it is
translated into a gate-level netlist.

\paragraph{\textbf{Seed Generator and Mutation Engine:}}  The seed generator, implemented using custom Python scripts, is designed to produce the initial seeds necessary for \synfuzz. The test suite generation is seamlessly integrated with both the seed generator and the feedback engine. Additionally, 
for the mutation engine, we follow the AFL mutation algorithm \cite{AFL'23}, deployed via Python scripts, and perform mutations based on the feedback provided by the feedback engine.

% \vspace{-1 em}

\subsection{Bugs Reported} \label{bugs}

We now present the synthesis bug detected by \synfuzz. \synfuzz~found seven new bugs in our selected IP benchmarks and ten existing bugs in open-source EDA synthesis tool Yosys \cite{SymbiYosys} as outlined in Table \ref{tb1:bugreport} and Table \ref{yosysbugs} respectively. Following this, we present the exploitation of one of these bugs via EDA tools to inject vulnerabilities in Section \ref{exploits}. Later in Section \ref{clima}, we present how a malicious library exploits the EDA synthesis to foster malicious library mapping. The CVEs assigned for the bugs reported by \synfuzz~  are provided in Appendix \ref{cve} .

\subsubsection{\textbf{Bugs in RSA}}
\begin{figure}[htb!]
\vspace{-1em}
  \centering
  \includegraphics[width=0.7\linewidth]{
  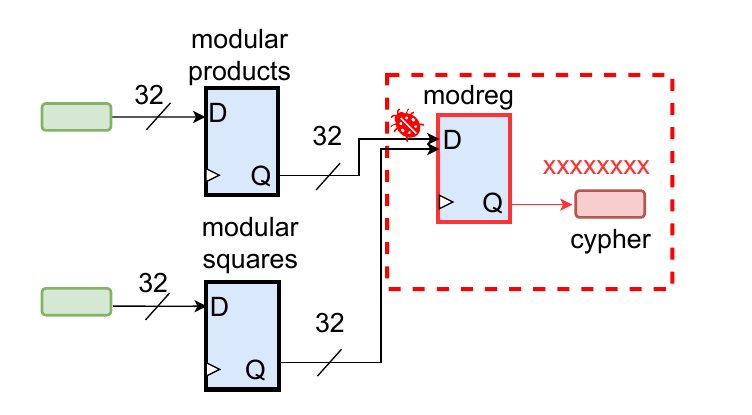} \vspace{-1em}
  \caption{Multiple Driver bug in the RSA}
 
   \label{fig:rsa}
  \vspace{-0.1em}
\end{figure}

\synfuzz~uncovered a critical bug in the synthesized netlist of the RSA design, where discrepancies were observed between the RTL and the gate-level netlist during the generation of ciphertext. 
The bug, \textbf{B1}, occurs in the RSA design, where it fails to generate ciphertext for any given input text. This issue arises from multiple sources driving the modular arithmetic operations register, which results in entering an undefined state, as shown in Figure \ref{fig:rsa}. The RSA performs multiple modular operations including products and squares as shown in Figure \ref{fig:rsa}. As illustrated in Figure \ref{fig:rsa}, we can see that both modular product and modular squares are connected to the same register  \verb|modreg|. Consequently, the modular operations, which are critical for the cryptographic computation and essential for generating ciphertext, become unavailable. This not only disrupts the RSA encryption process but also compromises the system's ability to perform secure and reliable cryptographic functions, leading to a denial of service (DoS) scenario for the affected hardware module. In Section \ref{exploits}, we successfully exploit this bug as a motivation to create a multiple driver in the DES crypto core through the EDA Synthesis Tools. This bug can be mapped to CWE-665 (Improper Initialization) or CWE-1419 ( Incorrect Initialization of Resource), resulting in unexpected behavior or security issues.

\subsubsection{\textbf{Bugs in PicoRV32}} 

The bug \textbf{B2}, resides within the processor's memory interface, where the address calculation for subsequent memory operations is not accurately updated. This flaw can result in disruptions to memory access, such as incorrect instruction fetching or inadvertent data retrieval from unintended memory locations, thereby compromising the reliability and integrity of the processor's operation. This bug is analogous to CWE-131 (Incorrect Calculation of Buffer Size) or CWE-119 (Improper Restriction of Operations within the Bounds of a Memory Buffer), which are typically considered from a software perspective but it is being applied in a hardware context.

\textbf{Bug B3:} The PicoRV32 processor includes an optional interface for integrating a custom co-processor to enable hardware acceleration. However, a bug has been identified in the Pico Co-Processor Interface, where instructions meant for the co-processor are not fully fetched into the decode stage, leading to incomplete or incorrect data being forwarded to the co-processor. This bug \textbf{B3} undermines the reliability of custom instruction execution, limiting the effectiveness and compatibility of hardware acceleration through the co-processor interface. This is similar to the bug described in CWE-1422 (Exposure of Sensitive Information caused by Incorrect Data Forwarding during Transient Execution).

\subsubsection{\textbf{Bugs in MIPS}} The bug \textbf{B4} resides within the Arithmetic Logic Unit (ALU) of the MIPS processor. The final output of the ALU operations fails to update when the computed result of the addition operation undergoes a change. As a result, any computational changes in the sum register do not affect the final output of the ALU,  causing the ALU to erroneously retain stale data. Consequently, this will lead to incorrect arithmetic operations. This bug can be mapped to similar CWEs, such as CWE-682 (Incorrect Calculation), or it may lead to issues described in CWE-190 (Integer Overflow or Wraparound).

\begin{table*}[h!]

	\centering
	\caption{ {Detection of synthesis translation bugs in open-source Yosys EDA Tool by \synfuzz} }
	\vspace{-1em}
	\label{yosysbugs}
	\scalebox{0.85}{
		
		\begin{tabular}{|c|l|c|}
			\hline  
			\textbf{Bug Reference\#\blzs} & \textbf{Bug Description} &  \textbf{Detected by \synfuzz~?}\\ \hline
			
			5105 &  \textbf{B8:} Incorrect optimization with Large Constant Shift & \cmark   \\
			
			\hline
            5099 &  \textbf{B9:} Incorrect shift optimization & \cmark \\
			
			\hline

            4491 &  \textbf{B10:} Custom Yosys Passes Result in Faulty Synthesis and Simulation Errors   & \cmark \\
			
			\hline
		4395 &   \textbf{B11:} Incorrect bit operation handling on empty strings & \cmark \\
			
			\hline
		4164 &  \textbf{B12:} Misoptimization of wide shifts bug & \cmark \\
			
			\hline
		4151 &  \textbf{B13:} Misoptimization of MUX Tree & \cmark \\
			
			\hline
		3895 &  \textbf{B14:} Inconsistency Issue with Continuous Assignment Error after FSM Optimization  & \cmark \\
			
			\hline
		2969 &  \textbf{B15:} The optimization of for-loop doesn't produce the expected behavior & \cmark \\
			
			\hline
		
			2648 &  \textbf{B16:} The unextend option incorrectly handle sign bits & \cmark \\
			
			\hline
		1161 &  \textbf{B17:} Multiple Drivers changes the functionality after optimization & \cmark \\
			
			\hline

			\hline
		\end{tabular}
}
\scriptsize

\blzs~Bug Reference\# refers to the Git Issue ID in the Yosys repository \cite{Yosys}. \\

\end{table*}
\subsubsection{\textbf{Bugs in or1200 Processor}}  The bug \textbf{B5} is in the front-end design of the or1200 processor between the Load Store Unit (LSU) and Debug Unit. The debugging unit, when performing debugging executions, requests a memory load operation. Upon receiving this request, the Load-Store Unit (LSU) retrieves the requested data from memory and makes it available to the debugging unit during the debugging process. It was detected that an incomplete data retrieval was performed from the LSU for the debugging request. This will lead to incorrect data received by the debugger leading to misinterpretation of memory states and impeding the debugging entity from diagnosing the issues in the processor. This bug can be generalized through relevant CWEs such as CWE-125 (Out-of-bounds Read) or CWE-119 (Improper Restriction of Operations within the Bounds of a Memory Buffer), which can lead to unpredictable behavior or security risks.

\textbf{Bug B6} arises when the instruction 95bf022d fails to execute in the synthesized netlist, resulting in an incomplete program counter (PC) update and undefined program flow. The instruction is executed, and the program counter fields are in an undefined state. This discrepancy causes disruptions in the program executions. Bug B6 can be considered analogous to CWE-451 (User Interface Misrepresentation of Critical Information), which, if not addressed, may result in incorrect execution or unintended consequences.

\textbf{Bug B7 } is the inaccurate update of program counter (PC) values when the values in the Next Program Counter (NPC) stored in the Special Purpose Registers (SPRs) change. The program counter is staled at the previous value when the value of NPC updates. This introduces the vulnerability of incorrect fetching of programs leading the processor to stall. Bug B7 may not have a direct mapping to a specific CWE, but it leads to issues that can be mapped to similar CWEs, such as CWE-221 (Information Loss or Omission) or CWE-664 (Improper Control of a Resource Through its Lifetime).

% \textcolor{blue}{
\subsubsection{\textbf{Bugs in Yosys EDA Tool}} Our proposed \synfuzz~also discovered ten existing synthesis translation bugs \textbf{(B8-B17)} in the open-source EDA synthesis tool Yosys. To demonstrate the efficacy of our \synfuzz~in detecting translation bugs introduced by EDA tools, we curated a corpus of existing bugs reported through public bug trackers \cite{Yosys}. Leveraging this dataset \footnote{The RTL designs provided to \synfuzz~were sourced from publicly available code snippets listed in the bug trackers of the Yosys EDA tool.}, we applied our proposed comprehensive fuzzing methodology—Inter-Tool, Intra-Library, Inter-Tool, and Inter-Library fuzzing strategies—as outlined in Section~\ref{difflib}. Our analysis revealed that the netlists generated by commercial EDA tools such as Synopsys Design Compiler and Cadence Genus consistently adhered to the RTL-level functional specification. In contrast, the netlists produced by the open-source tool Yosys exhibited translation-induced functional bugs. The existing bugs in the open-source Yosys EDA tool detected by \synfuzz~are detailed in Table~\ref{yosysbugs}, along with their corresponding bug reference numbers as reported in publicly available bug trackers.
% }

\subsection{Bug Exploitation} \label{exploits}

Based on the detection of multiple driver bugs in RSA, we intentionally replicate and introduce similar vulnerabilities into one of our benchmarks, the Data Encryption Standard (DES), by exploiting EDA tools. This approach demonstrates how EDA tools can be leveraged to introduce malicious modifications into hardware designs.

\paragraph{\textbf{Assumption:}}For this purpose, we assume the role of a malicious hardware developer who exploits the state-of-the-art features of an EDA tool to transform a benign piece of hardware into a compromised or malicious component.

\paragraph{{\textbf{Multiple Driver Exploit:}}}In the context of our findings from \textbf{Bug B1}, the creation of multiple-driver bugs involves introducing conditions where library-mapped cell gates are driven by multiple sources, leading to undefined or incorrect behavior. To successfully introduce such vulnerabilities, an attacker must first conduct a thorough analysis of the design to identify critical registers and computational components that are integral to the functionality of the IP. These components serve as prime targets for manipulation due to their significance in the operational integrity of the design.

In our specific exploit, we focused on the substitution box (S-box) of DES, which is a crucial computational block responsible for the non-linear substitution of data. The S-box is essential for ensuring the security of the encryption process, making it an ideal target for introducing a vulnerability. Once the critical components, such as the S-box, have been identified, an attacker can utilize advanced EDA tool features to deliberately create a multiple-driver scenario within the design.

This process involves deploying sophisticated EDA tool commands to establish multiple connections to the targeted components, effectively creating conflicts in the driver-load connectivity. The attacker can specify the desired drivers and loads within the tool, intentionally assigning multiple sources to drive the same gate or node. By doing so, the attacker introduces ambiguity in the design’s behavior, which could lead to undefined outputs, functional failures, or exploitable vulnerabilities during operation.
This exploitation highlights the risks of trusted EDA tools being misused to introduce subtle vulnerabilities, emphasizing the need for robust security, thorough validation, and secure-by-design principles in modern hardware workflows.

% \vspace{-1 em}
\subsection{Coverage Analysis}
In this section, we present the coverage reports for the benchmarks utilized by our fuzzer. For relatively smaller designs, such as UART, GPIO, DES, MIPS, RSA, and DSP, the fuzzer is executed for up to 8 hours. If the design achieves 100\% coverage before this time, the fuzzer is terminated early. For larger designs, including or1200, IBEX, MIPS, and PicoRV32, the experiments are conducted for 12 hours to ensure comprehensive fuzzing. As mentioned in Section \ref{libtog}, we extract the library coverage for our designs.

\begin{figure}[htb!]
\vspace{-1em}
  \centering
  \includegraphics[width=1\linewidth]{
  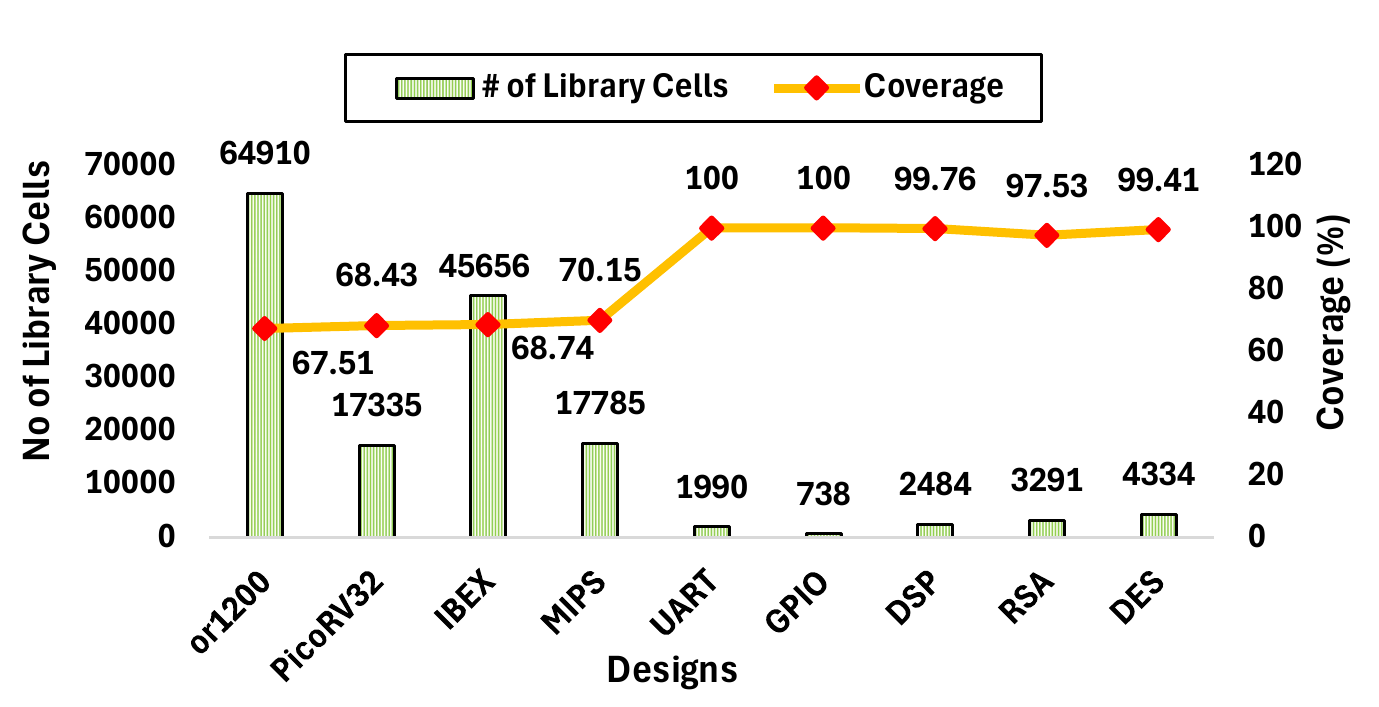} \vspace{-1em}
  \caption{Coverage Results for benchmarks in \synfuzz}

   \label{fig:coverage}
  % \vspace{-1em}
\end{figure}

The coverage performance evaluation of \synfuzz~highlights the effectiveness and adaptability of the fuzzer across a diverse range of designs. As outlined in Figure \ref{fig:coverage}, or1200 is the largest design with 64190 library cells. For smaller designs such as UART, GPIO, DSP, RSA, and DES, the fuzzer achieves exceptional coverage, close to 99\%.  For larger designs like or1200, PicoRV32, and IBEX, with higher library cell counts, the fuzzer achieves notable coverage values of 
67.51 \%, 68.43 \%, and 70.15\%. These results are in line with some of the existing works \cite{Kande'22,hybrid}, and underline the proposed fuzzer’s capability to handle complex architectures effectively, even within a limited fuzzing duration. On Overall, the results affirm the fuzzer’s robustness and scalability, excelling in smaller designs and performing admirably on larger ones, making it a versatile tool for comprehensive synthesis bug detection.

\begin{figure}[htb!]
% \vspace{-1em}
  \centering
  \includegraphics[width=1\linewidth]{
  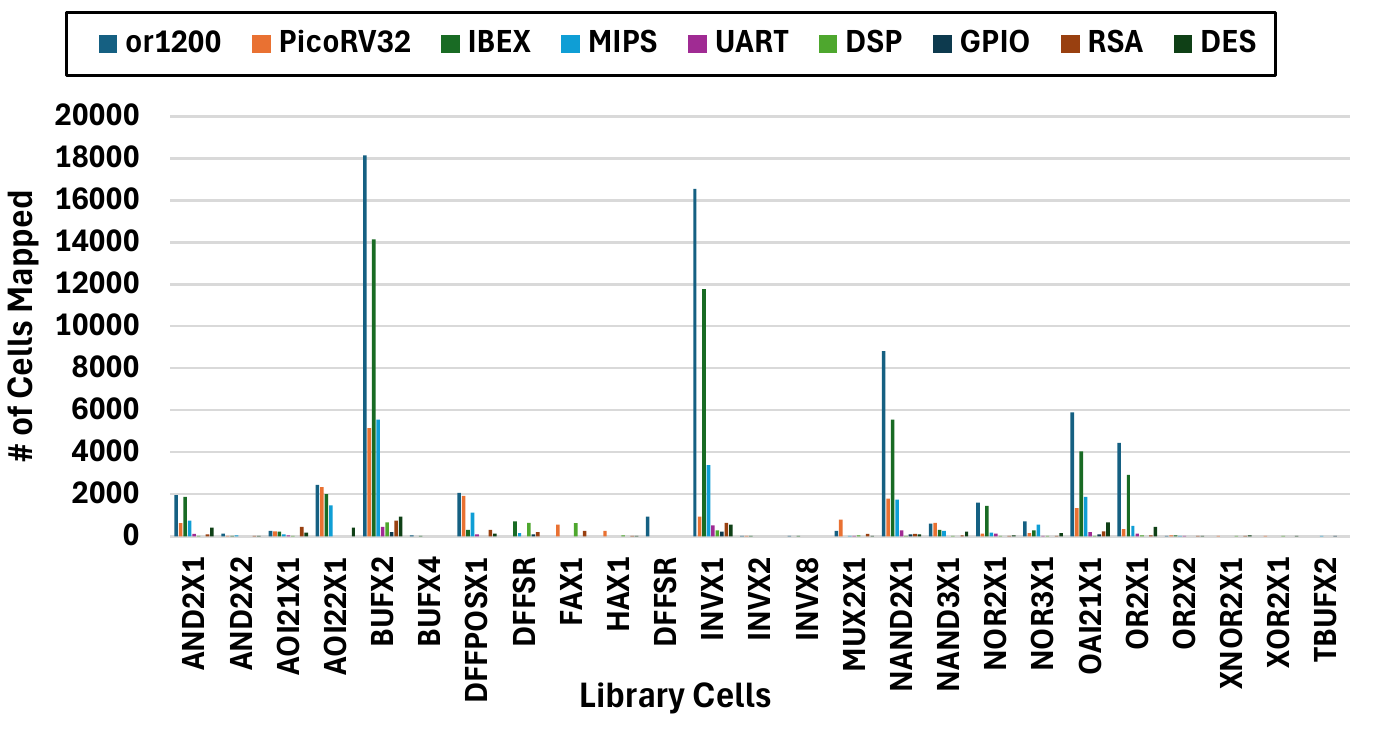} \vspace{-1em}
  \caption{Cell Mapping utilized by our benchmarks}
 
   \label{fig:cell}
  \vspace{-0.1em}
\end{figure}

Furthermore, we present the utilization of cell mapping as applied to our benchmarks in Figure \ref{fig:cell}. Our complex and diverse set of hardware designs effectively maps to approximately 91.70\% of the cells available in the library. This high mapping percentage underscores our intent to thoroughly test the library cells under various conditions to evaluate their performance and robustness. Among the available cells, OR2X2, XNOR2X1, and XOR2X1, as well as TBUFX2, are the least utilized mapping cells. In contrast, INVX1, BUFX2, and NAND2X1 are the most highly utilized cells, reflecting their critical role and frequent occurrence in our designs.

% \vspace{-1 em}
% \subsection{Comparison with Conformal}
\subsection{Comparison with Logic Equivalence Checking}\label{leceval}

We also compare our \synfuzz~with another industry standard approach - Logic Equivalence Check (LEC)\cite{LEC_cadence}, which is a popular technique used for the formal verification of hardware designs. During LEC, the structural and functional features of the synthesized design are compared against the RTL as a reference model after the synthesis to determine whether the two designs exhibit the same behavior. The semiconductor industry relies on commercial EDA tools such as Cadence Conformal \cite{Conformal} for LEC. LEC operates by mapping and comparing specific structural and functional points, such as inputs, outputs, and intermediate signals, between the synthesized design and the RTL. These mapped comparison points are critical for identifying any discrepancies and ensuring that both designs exhibit the same behavior.

However, there are two challenges in performing these comparisons. Firstly, as modern hardware designs grow increasingly complex and larger in scale, the mapped comparison points increase exponentially; consequently, the complexity of LEC rises. This can increase the verification period to map the comparison points and verification period producing ambiguous results. Secondly, conventional LEC-based verification can perform poorly (or fail) under the following scenarios: structural optimizations introduced by EDA tools during synthesis and P\&R under strict design constraints; technology translation from one standard cell library to another having different process corners or missing logic functions; untracked port/net/gate renaming introduced by EDA tools or designers; and functional/structural design transformations for IP protection like state-space transformation \cite{pip_2023}. 

To ensure proper evaluations, the translated information must be accurately fed into the LEC tool, a process that demands significant human expertise and effort. The intricate nature of the mapping and comparison points, coupled with the nuances of design transformations and tool-specific optimizations, necessitates a deep understanding of both the design and the LEC tool’s operation, hindering the efficiency in bug detection. The LEC fails to detect bugs found by our \synfuzz. Thus, in contrast to \synfuzz, the existing formal tools are not scalable for large designs, are prone to errors, and require significant human expertise to operate effectively.
% \vspace{-1 em}

\section{CLiMA: Compromised Library Mapping Attack} \label{clima} 

In this section, we present our Compromised Library Mapping Attack (CLiMA), which exploits the optimizations of EDA synthesis tools to produce a malicious version of the hardware design. We are familiar with the fact that during the synthesis process, library binaries provided by fabrication vendors are fed to the EDA synthesis tool to map the design to library gate cells. However, an adversary or a malicious library vendor can exploit the optimizations of the EDA tool to manipulate the mapping process and create a malicious version of the hardware. To show the practicality of this attack, we developed a malicious version of an open-source 45 nm library and used it to create a malicious variant of the or1200 CPU design. The CLiMA attack is non-detectable by traditional formal hardware verification methods such as LEC. Based on the malicious mapping process there exist two scenarios \textit{\textbf{ 1) Random Mapping Exploitation 2) Targeted Mapping Exploitation.}} Before we present these attacks we provide some background on the library binary and the EDA tool optimizations which are being exploited.

\textbf{Background:} The library file (.lib) contains comprehensive information about the available gate cells, including their respective parameters such as functionality, area, power consumption, and timing characteristics. This detailed description serves as the foundation for mapping the RTL design to the appropriate standard-cell gates during the synthesis process. These are compiled by the library compilers which often emit the provided library in a binary format (.db). Since these binaries are not readable any alterations made are non traceable. A malicious vendor can manipulate the library file parameters (.lib) before it is compiled into a binary file (.db), masking the abnormalities. This tampered binary file (.db) when used by the synthesis tool, can lead to incorrect functionality or security vulnerabilities in the final design. After this, when simulating the netlist with the library environment (usually in .v), the tampered binary file causes incorrect gate mapping resulting in a functional mismatch.

To successfully execute this attack, the attacker must analyze the parameters of the library file to identify which attributes need to be tampered with to make it appear more attractive to the EDA tool. This involves understanding how the EDA tool prioritizes gate selection during synthesis based on characteristics such as functionality, timing, power, and area. By carefully manipulating these parameters, the attacker can influence the synthesis process to favor the maliciously modified gate cells, embedding vulnerabilities or altering the design’s intended behavior without detection. In addition, there are design constraint commands such as maximum/minimum area, and power which influence the synthesis process. In our attack, we target for optimizing the synthesis process with minimum area constraint. In our malicious version of 45nm, we manipulate the functionality and area of certain cells to make it attractive for the EDA tool during the synthesis process. However, one can also do the same by exploiting minimum dynamic power and leakage power parameter constraints.

\begin{table}[htbp!]
	\centering
	\caption{Instance Mapping Before and After CLiMA}
	\vspace{-1 em}
	\begin{tabular}{|c|c|c|c|c|}
		\hline
		\textbf{Library Cells} & \multicolumn{2}{|c|}{\textbf{Original}} & \multicolumn{2}{|c|}{\textbf{After CLiMA}} \\ \hline
		& \textbf{\#} & \textbf{Area}    & \textbf{\#} & \textbf{Area}    \\ \cline{2-5}
		AND2X1           &   384        & 901.05                 & 598          & 805.20                \\ \hline
		AND2x2  & -& -&    355         & 999.60             \\ \hline
		
		A0121X1           & 39          & 109.81               & 65           & 183.02                    \\ \hline
		A0122X1          & 556           & 1826.51                & 557           & 1829.80                 \\ \hline
		BUFX2           & 1662           & 3899.88                & 1627           & 3817.75                 \\ \hline
		INVX1           & 2005          & 2822.84                & 2008           & 2827.06               \\ \hline
		NAND2X1           & 725           & 1360.97               & 750           & 1407.90                   \\ \hline
		NAND3X1          & 105          & 246.38                    & 55           & 129.05           \\ \hline
		NOR2X1          & 361          & 847.08                   & 324           & 760.26                  \\ \hline
		NOR3X1           & 89           & 250.60                  & 55           & 154.86              \\ \hline
		OAI21X1           & 218           & 613.84                  & 216           & 608.21                   \\ \hline
		OR2X1         & 439       & 1030.11                 & -        & -               \\ \hline
		OR2X2           & 2           & 5.63                  & 1           & 2.81                \\ \hline
		Total          & 6585          & 13914                & 6611           & 13525.58                  \\ \hline
	\end{tabular}
	% \vspace{-1em}
	\label{tab:clima}
\end{table}

% \vspace{-1 em}
\subsection{Random Mapping Exploitation} \label{random}For implementing this attack, we selected the ALU  of the or1200 processor as our target module. The ALU is a critical component responsible for performing arithmetic and logical operations, making it an ideal candidate for demonstrating the potential impact of a malicious library mapping attack. By focussing on the ALU, we demonstrate how a tampered library will lead to incorrect arithmetic operations. In our malicious version of 45nm library, the targeted cell is an AND2X1 gate 
where 2X1 denotes a two-input AND gate with a standard drive strength of X1. The functionality of the AND2X1 gate in the malicious version of the library is modified to perform an OR logic operation instead. Additionally, the area of these cells is intentionally reduced to be smaller than the area of the existing OR gate cells (OR2X1, OR2X2) in the library. As a result, the synthesis tool prioritizes mapping the AND2X1 cells for implementing OR logic, aiming to achieve the minimum area requirements specified during the optimization process. This malicious version of the library is compiled with library compilers and passed on to the EDA tools for the synthesis process.

As outlined in Table \ref{tab:clima}, a significant reduction in the design area can be observed before and after the execution of the library mapping attack. Post-attack, there are no OR2X1 cells mapped in the design, as the malicious AND2X1 cells, modified to implement OR logic, appear more attractive to the EDA tool for meeting minimum area requirements. Specifically, 439 OR2X1 cells were replaced with AND2X1 cells due to the CLiMA attack. When simulating the netlist with the library environment, 598 cells implement AND functionality instead of OR  functionality using the malicious AND2X1 cells, resulting in an erroneous change in the ALU's functionality. Hence, the root cause for this incorrect logic implementation is hard to find as this is masked in the malicious library. 

%\vspace{-1 em}

\begin{table}[htbp!]
	
	\centering
	\caption{ {Design parameter deviations before and after the targeted mapping attack} }
	\vspace{-1em}
	\label{tab:target}
	\scalebox{0.85}{
		
		\begin{tabular}{|c|c|c|c|}
			\hline  
			\textbf{} & \textbf{Instances} &  \textbf{Area ($\mu m^2$)} & \textbf{Power ($\mu W$)}  \\ \hline
			
			Before &  1056 & 2783.59 &  179 \\
			
			\hline
			After &   1056  & +0 \% &  + 0 \% \\

			\hline
		\end{tabular}
	}
    \vspace{-1em}
\end{table}

\begin{table*}[!ht]
	
	\centering
	\caption{Comparison with existing HW Fuzzing Frameworks  }
	\vspace{-1em}
	\label{table:prior}
	\scalebox{0.75}{
		\begin{tabular}{|l|l|l|l|l|l|c|}
			% \toprule
			\hline
			\textbf{Fuzzer} & \textbf{Targeted Abstraction} & \textbf{Input} & \textbf{Simulator} & \textbf{Coverage Metric} & \textbf{Target Design } & \textbf{Synthesis Bugs}    \\
			\hline
			
			TheHuzz \cite{Kande'22}  & RTL (HW)  & Assembly  & Synopsys VCS & \begin{tabular}[c]{@{}l@{}} FSM, Branch, toggle, \\ conditional \end{tabular}  & RISC-V  & \textcolor{red}{\xmark}  \\
			\hline
			
			Processor Fuzz \cite{Canakci'23} & RTL (HW) & Assembly & Verilator & \begin{tabular}[c]{@{}l@{}}Control path register, \\ ISA-transition \end{tabular} & RISC-V  & \textcolor{red}{\xmark} \\   
			\hline
			
			RFUZZ \cite{Laeufer'18}  & FIRRTL  & Series of bits & Any & Mux Toggle & Peripherals, RISC-V & \textcolor{red}{\xmark}  \\
			\hline
			DifuzzRTL \cite{Hur'21}  &  RTL (SW)  & Assembly & Any & Register Coverage & RISC-V CPU & \textcolor{red}{\xmark}  \\
			\hline
			Trippel et al \cite{Trippel'22} & RTL (SW)  & Byte Sequence & Verilator & Edge Coverage & AES,HMAC,KMAC, Timer  & \textcolor{red}{\xmark}  \\
			\hline
			HyperFuzzer \cite{hyperprop}  & RTL (SW) & Series of bits & Verilator & High-level & SoC  & \textcolor{red}{\xmark} \\
			\hline

			SoCFuzzer \cite{Hossain'23}  &  RTL (HW)  & Byte Sequence & Xilinx ISA & \begin{tabular}[c]{@{}l@{}}Randomness, target output, \\ input coverage \end{tabular} & SoC & \textcolor{red}{\xmark}  \\
			\hline
			HyPFuzz \cite{hybrid} &  RTL (HW)  & \begin{tabular}[c]{@{}l@{}} Assertion Cover \\ Properties, \\ Byte Sequence \end{tabular}  & \begin{tabular}[c]{@{}l@{}} Jasper Gold, \\ Synopsys VCS \end{tabular}  & \begin{tabular}[c]{@{}l@{}} FSM, Branch, toggle, \\ conditional \\ \end{tabular}  & RISC-V  & \textcolor{red}{\xmark}  \\
			% && Byte Sequence & Synopsys VCS&conditional & &  \\
			\hline
			
			\textbf{\textit{SynFuzz (This work)}} & \begin{tabular}[c]{@{}l@{}} \textbf{Library mapped} \\ \textbf{gate-level netlist} \end{tabular} & \textbf{Series of bits} & \begin{tabular}[c]{@{}l@{}} \textbf{Cadence Xcelium} \textbf{} \end{tabular} & \begin{tabular}[c]{@{}l@{}} \textbf{Library Toggle Coverage,} \\ \textbf{Expression} \end{tabular} & \begin{tabular}[c]{@{}l@{}} \textbf{CPU Designs, RSA, DES,} \\ \textbf{UART, GPIO, DSP} \end{tabular} & \textbf{\cmark} \\
			
			\hline

			%References & 8pt, no page limit, list \\
			%          & all authors' names\\
			%\hline
			
		\end{tabular}
	}
\end{table*}

\subsection{Targeted Mapping Exploitation} \label{targetclima} 
\begin{figure}[htb!]
	\vspace{-1em}
	\centering
	\includegraphics[width=0.7\linewidth]{
		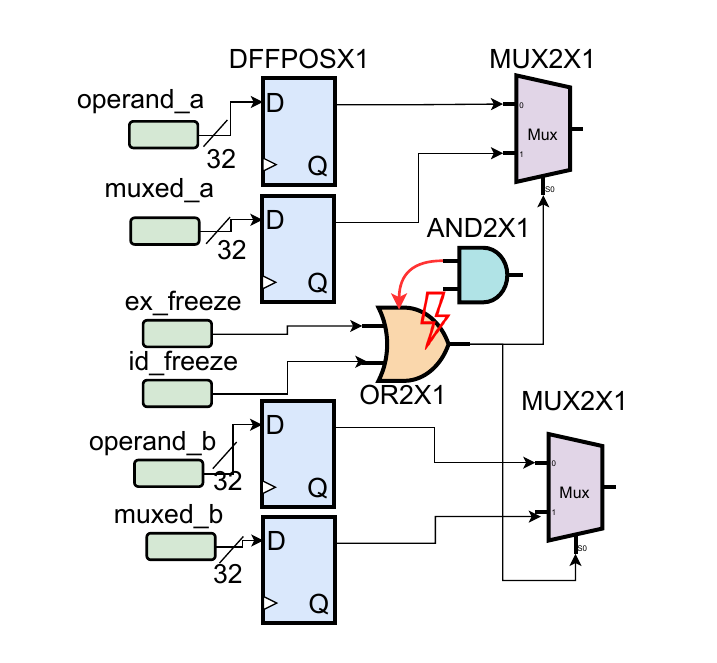} \vspace{-1em}
	\caption{Targeted Library Mapping Attack in or1200 processor
		%\textcolor{red}{You have white space in fig, you need to remove it.}
		% \textcolor{blue}{acknowledged}
		% \cite{9} \textcolor{red}{Did you redraw this figure?\textcolor{violet}{Done}}
	} \label{fig:wb}
	% \vspace{-0.5em}
\end{figure}

In this attack, instead of randomly mapping cells, specific cells are deliberately targeted to create malicious behavior as shown in Figure \ref{fig:wb}. By modifying the functionality or characteristics of these chosen cells, the attacker ensures that the synthesis tool prioritizes their use in critical parts of the design, embedding vulnerabilities or altering functionality in a controlled and intentional manner. For implementing this attack, we specifically targeted the operand multiplexer module, modifying its behavior to introduce incorrect register forwarding of data. This deliberate manipulation ensures that the synthesized design mishandles data flow between registers, disrupting the forwarding to computing modules. Table \ref{tab:target} presents the number of cell instances and the incurred area and power overhead before and after the targeted mapping attack. The results clearly demonstrate that the attack does not introduce any additional overhead, making it practically viable with no adverse impact on the performance of the design.

%\vspace{-1 em}

\section{Related Works}

In this section, we outline the limitations of existing works in Table \ref{table:prior} and demonstrate how \synfuzz~effectively detects synthesis bugs.

\begin{formal}
\textbf{Observation O2: Comparing our work with previous approaches is not directly suitable due to inherent differences in abstraction levels and coverage metrics. The methodologies and evaluation criteria of prior works focus on specific metrics and levels of abstraction that are not aligned with the scope and objectives of our chosen level of abstraction, input format, coverage, and reference model. As such, a direct comparison would not yield meaningful insights and may lead to inaccurate and false interpretations. }
\end{formal}

The majority of popular fuzzing frameworks, such as TheHuzz \cite{Kande'22}, DifuzzRTL \cite{Hur'21}, and ProcessorFuzz \cite{Canakci'23} are CPU design fuzzers, operating at the RTL level of abstraction and typically compare their outputs with ISA simulators, such as Spike simulator. However, the chosen level of abstraction and the input strategies used in these existing works are not efficient in uncovering synthesis bugs.

Prior works, as summarized in Table \ref{table:prior}, have explored fuzzing hardware at different levels of abstraction. For instance, RFuzz \cite{Laeufer'18} was an early effort aimed at fuzzing RTL designs by directly adopting software fuzzers and using metrics like multiplexer toggles for coverage. However, it fails to scale effectively and often misses bugs in complex designs. In contrast, Trippel \textit{et al.} \cite{Trippel'22} proposed fuzzing hardware-like software, rather than directly applying software fuzzers to hardware designs. While this approach holds promise, the translated hardware models do not support HDL constructs or accurately account for intrinsic hardware characteristics. Moreover, the coverage metrics used in these approaches are not suitable for hardware abstractions.

Additionally, works such as \cite{hyperprop} proposed leveraging hyperproperties—
higher-level properties describing security policies by comparing system behaviors. Although useful for certain applications, these methods require significant human intervention and deep design knowledge, limiting their scalability for complex SoC designs. Later efforts \cite{Kande'22,hybrid,Canakci'23} focused on fuzzing hardware at its native hardware abstraction level, as detailed in Table \ref{table:prior}. These frameworks were designed to account for the unique characteristics of hardware systems, making them efficient at detecting bugs. However, they still rely on coverage metrics aligned with the traditional IC design flow and remain restricted to higher levels of abstraction. Based on the motivation of compiler fuzzing from software, TransFuzz \cite{solt_mirtl_2025} is proposed to fuzz the tools and compare the performance with Verismith, an open-source hardware tool fuzzer, but never focus on synthesis bugs introduced in the design.

It is important to note that a direct comparison between our work and these prior approaches is unsuitable. The abstraction levels, input strategies, and coverage evaluation metrics used in these existing works differ significantly from ours and do not apply to gate-level netlist abstraction. Notably, none of the prior frameworks focus on fuzzing at the gate-level netlist abstraction, which is the core of our methodology for identifying synthesis bugs, library bugs, and their vulnerabilities.

In contrast, our proposed \synfuzz~1) supports conventional hardware design and verification methods, 2) captures intrinsic hardware characteristics at gate-level netlist, 3) is efficient in detecting bugs and vulnerabilities associated with the synthesis and malicious library vendors, 4) does not require extensive design knowledge or expertise in hardware design, and 5) is scalable large to designs with several thousands of library cells.

% \vspace{-0.75em}

%\vspace{-1 em}
\section{Discussion and Limitations}
\label{discussion}

\paragraph{\textbf{RTL and Netlist Accessibility:}} \synfuzz~relies on access to RTL designs and EDA tools to effectively detect synthesis bugs. Verification engineers typically have access to these RTL codes during the design and verification process, enabling them to simulate and analyze the design behavior. Additionally, RTL codes can be acquired from third-party vendors at a cost, providing flexibility for organizations or engineers who require pre-designed or specialized modules for their projects. This accessibility facilitates the use of \synfuzz~to validate both custom and third-party designs for potential synthesis vulnerabilities.

\paragraph{\textbf{FPGA Synthesis Bugs:}} \synfuzz~can be extended to fuzz FPGA-based frameworks to detect synthesis bugs in FPGA synthesis tools such as Xilinx Vivado \cite{vivado} and others. In FPGA designs, device-specific library files, often referred to as bitstream libraries or technology libraries, define the mapping of high-level designs to FPGA-specific resources like Look-Up Tables (LUTs), Flip-Flops, and Block RAMs (BRAMs). These libraries play a critical role in determining the final implementation of the design on the FPGA. By targeting these device-specific libraries, \synfuzz~can identify bugs that arise due to incorrect mapping of RTL logic to FPGA primitives. One can also perform DiffLib in the context of device libraries. Extending \synfuzz~to FPGA frameworks allows for comprehensive validation of FPGA primitives ensuring robust and reliable hardware implementations on programmable platforms.

\paragraph{\textbf{Fuzzing Temporal Characteristics:}} As mentioned in Section \ref{setup}, all the simulations are under Zero Delay Simulation conditions ignoring all the timing intent. \synfuzz~is not suitable for detecting any timing bugs and vulnerabilities. However, one can extend \synfuzz~to detect timing vulnerabilities introduced by the library cells. 
% \vspace{- 1 em}

% \vspace{-1 em}
\section{Conclusion}
\label{conclusion}
Hardware bugs are increasingly prevalent at different levels of abstraction during various stages of modern IC design. Existing fuzzing frameworks fall short in detecting bugs associated with synthesis and library bugs.  To address this gap, we present our novel framework, \synfuzz, which focuses on fuzzing at the library mapped gate-level netlist. \synfuzz~demonstrates its efficiency by identifying 7 new synthesis bugs in popular open-source designs. We presented DiffLib to extensively fuzz the EDA library in different EDA tools to uncover library vulnerabilities. By strategically exploiting EDA tool optimization settings, we presented and created a compromised library mapping attack that is undetectable by LEC. While traditional LEC methods face limitations in identifying such issues, \synfuzz~overcomes these challenges, providing a more robust and comprehensive solution for synthesis bug detection. Compared to Conformal, a formal LEC verification tool, \synfuzz~eliminates the need for tool-specific and design knowledge while addressing other limitations.

\paragraph{\textbf{Responsible disclosure}} The bugs have been reported to respective design maintainers. 

\appendix
\section{CVE Allocation} \label{cve}

\begin{table}[htbp!]
\centering
\label{tab:tabcve}
	\caption{ {CVEs assigned for bugs found by \synfuzz} }
\begin{tabular}{|l|l|}
\hline
Bug \# & CVE \#     \\ \hline
B2, B3 & 2025-51678 \\ \hline
B5     & 2025-51677 \\ \hline
B6     & 2025-51679 \\ \hline
B7     & 2025-51675 \\ \hline
\end{tabular}
\end{table}

\bibliographystyle{ACM-Reference-Format}
\bibliography{ref,references,reference}

% \begin{acks}
% \end{acks}

% %%
% %% The next two lines define the bibliography style to be used, and
% %% the bibliography file.
% \bibliographystyle{ACM-Reference-Format}

% %%
% %% If your work has an appendix, this is the place to put it.
% \appendix









\end{document}